# Design Challenges for GDPR RegTech

Paul Ryan[a], Martin Crane[b] and Rob Brennan[c]
*Uniphar PLC, ADAPT Centre, School of Computing, Dublin City University, Glasnevin, Dublin 9, Ireland*
*paul.ryan76@mail.dcu.ie, {martin.crane, rob.brennan}@dcu.ie*

Abstract: The Accountability Principle of the GDPR requires that an organisation can demonstrate compliance with the regulations. A survey of GDPR compliance software solutions shows significant gaps in their ability to demonstrate compliance. In contrast, RegTech has recently brought great success to financial compliance, resulting in reduced risk, cost saving and enhanced financial regulatory compliance. It is shown that many GDPR solutions lack interoperability features such as standard APIs, meta-data or reports and they are not supported by published methodologies or evidence to support their validity or even utility. A proof of concept prototype was explored using a regulator based self-assessment checklist to establish if RegTech best practice could improve the demonstration of GDPR compliance. The application of a RegTech approach provides opportunities for demonstrable and validated GDPR compliance, notwithstanding the risk reductions and cost savings that RegTech can deliver. This paper demonstrates a RegTech approach to GDPR compliance can facilitate an organisation meeting its accountability obligations.

## 1 INTRODUCTION

In May 2018, the European Union (EU) introduced the GDPR. This regulation brought a high level of protection for data subjects, but also a high level of accountability for organisations (Buttarelli 2016). The GDPR principle of accountability requires that a data controller must be able to demonstrate their compliance with the regulation (GDPR Recital 74). This requires an organisation "to act in a responsible manner, to implement appropriate actions, to explain and justify actions, provide assurance and confidence to internal and external stakeholders that the organisation is doing the right thing and to remedy failures to act properly" (Felici, 2013).

Organisations can be complex entities, performing heterogeneous processing on large volumes of diverse personal data, potentially using outsourced partners or subsidiaries in distributed geographical locations and jurisdictions. A challenge to complying with the accountability principle of the GDPR for organisations is demonstrating that these complex activities and structures are meeting their regulatory obligations. The organisation must implement appropriate policies, procedures, tools and mechanisms to support their accountability practices (Felici, 2013).

Many organisations appoint a Data Protection Officer (DPO) to assist in this process. Bamberger describes the role as "the most important regulatory choice for institutionalising data protection" (Bamberger, 2015). In practice the DPO is the early warning indicator of adverse events when processing personal data within the organisation (Drewer, 2018). The DPO must have "professional qualities and, in particular, expert knowledge of data protection law and practices" (GDPR Art 37). This challenging role requires the DPO to monitor compliance and advise the organisation accordingly. The DPO acts independently of the organisation to assess and monitor the consistent application of the GDPR regulation and to ensure that the rights and freedoms of data subjects are not compromised (Article 8, EU charter). The role of DPO encompasses a dynamic motion of policy generation, staff training, business process mapping and review, compliance record keeping, audit, data protection impact assessments, and compliance consultations (Drewer, 2018). The constant pace of business change allied with evolving legal interpretations require constant vigilance on the part of the DPO and create additional challenges for accountability. Fundamentally, it is the organisation, and not the DPO, that must be able to demonstrate that it is meeting the threshold that is the accountability principle.

[a]  https://orcid.org/0000-0003-0770-2737
[b]  https://orcid.org/0000-0001-7598-3126
[c]  https://orcid.org/0000-0001-8236-362X

There are many solutions available to DPOs and organisations to help meet this challenge of demonstrating compliance to the accountability principle. This paper will evaluate the range of available tools, such as: privacy software solutions from private enterprise vendors, maturity models and regulator self- assessment tools. Despite the many GDPR compliance tools available, this paper will highlight that the majority fail to meet the accountability principle. Most are not supported by published methodologies or evidence for their validity or even utility. They lack the ability to integrate or be integrated with other tools and the level of automation and innovation in this space has also been limited.

In contrast, RegTech has emerged as a framework for automating regulatory compliance in the Financial Industry. The "Global Financial Crisis (GFC)" of 2008 prompted financial regulators to introduce new compliance regulations (Johansson, 2019), resulting in significant compliance challenges and compliance costs for organisations due to the complexity of these regulations. Strong data governance and mapping regulatory compliance provisions into software code (Bamberger, 2009) to facilitate regulatory compliance has been enabled by developments such as process automation, the digitising of data, the use of semantic methods and machine learning algorithms. RegTech uses such tools to efficiently deliver compliance and risk reports in integrated toolchains. The evolution of RegTech has shown that information technology can be used to support automated or semi-automated regulatory monitoring and reporting of compliance (Arner, 2017).

This paper proposes challenges for realising a RegTech approach to GDPR compliance whereby organisations leverage modern information technology to improve the organisational and external visibility of their GDPR compliance level. This approach requires automated data collection from relevant sources throughout the organisation and monitoring via GDPR compliance evaluation functions that could provide interoperable and machine-readable compliance metrics or reports for the organisation, suggested compliance actions and root cause analysis of compliance issues, using agreed data quality standards such as ISO8000.

The role of monitoring, analysing and reporting the GDPR compliance status in an organisation is the task of the DPO. A RegTech approach to GDPR compliance could provide the DPO with the ability to track organisational compliance progress, identify areas of compliance weakness and benchmark their performance against other organisations. This would greatly enhance an organisation's ability to demonstrate and improve compliance and thus meet the GDPR accountability requirement.

Section 2 will discuss the accountability principle and what it means in practice to an organisation and the challenges they face to meet the accountability principle. The role of the DPO, and their part in compliance will be discussed in detail from the perspective of a practising DPO. Section 3 reviews the current approaches to GDPR compliance and critiques the many available offerings such as private enterprise software solutions, maturity models and self-assessment checklists. Section 4 examines the financial Industry to see how RegTech is enhancing compliance using data driven solutions. Section 5 describes the challenges that must be faced in developing the next generation of GDPR compliance tools based on RegTech and documents the requirements that a DPO would require in such tools. Section 6 will introduce a proof of concept where a Data Protection Regulators self- assessment checklist has been utilised based on RegTech best practice, to provide a simple efficient method to demonstrate GDPR compliance and meet the requirements of the accountability principle.

## 2 THE GDPR ACCOUNTABILITY - A VIEW FROM THE DPO

In this section, this paper will discuss what the accountability principle of the GDPR means to organisations. The paper will look at the challenges that organisations are facing with demonstrating that they are meeting these obligations and it will discuss the role of the DPO in this process.

The Anglo-Saxon word "Accountability" has a broadly understood meaning of how responsibility is exercised and how it is made verifiable (Article 29 Working Party, 2010). Accountability can be viewed to be an expression of how an organisation displays "a sense of responsibility—a willingness to act in a transparent, fair and equitable way" (Boven's, 2007) and "the obligation to explain and justify conduct' (Boven's, 2007). The GDPR accountability principle requires a data controller "implement appropriate and effective measures to put into effect the principles and obligations of the GDPR and demonstrate on request" (Article 29 Working Party, 2010). In 2018 the Centre for Information Policy Leadership (CIPL) developed accountability-based data privacy and governance programs to encompass the key elements of accountability as described in Fig 1.

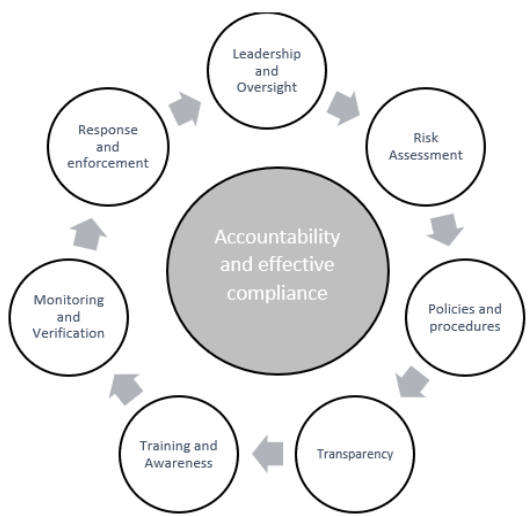

Figure 1: The Accountability Wheel– Universal Elements of Accountability (CIPL, 2018).

In practice, this can be viewed as "setting privacy protection goals based on criteria established in law, self-regulation and best practices and vesting the organisation with the responsibility to determine appropriate, effective measures to reach these goals" (CIPL, 2018). This is quite a challenging task for a data controller when you are dealing with a substantial legal text like the GDPR. There is a "lack of awareness of their obligations and duties in relation to personal data protection, it is urgent to define a methodology to be able to comply with the GDPR" (Da Conceicao Freitas, 2018).

In theory, the GDPR provides for certification methods in article 42 and 43 of the GDPR to assist a controller in demonstrating compliance. However, in practice this has proven to be a challenge for organisations as the European Union has not approved any Certification body to certify compliance (Lachaud 2016). In fact, there are views being expressed that the GDPR certification process cannot be successful. (Lachaud,2016).

Many organisations appoint a DPO to assist with their GDPR compliance, however it is important to note that the demonstration of compliance obligations ultimately rests with the controller (organisation) and not the DPO. The role of DPO within the organisation covers a wide range of tasks as prescribed in Article 39 of the GDPR. The main tasks are to monitor, inform and advise the controller or processor regarding compliance with the GDPR, to provide advice such as data protection impact assessments, to provide training and awareness raising and to co-operate with and act as a contact point for the supervisory authority.

The role of DPO requires a broad set of skills in GDPR legal compliance, and a detailed knowledge of business processes (Drewer,2018). The DPO works with numerous stakeholders such as data subjects, employees, processors and regulators and provides consultancy and guidance on business processes. The role involves a broad spectrum of activities from maintaining a register of processing activities to dealing with data breaches, to completing data protection impact assessments. The DPO must have visibility of all activities and monitor and report compliance to the highest level in the organisation (see Fig.2). The DPO is in essence "privacy on the ground" (Heimes, 2016), in that the DPO is the early warning system for GDPR compliance within the organisation (Drewer, 2018). The challenge for the DPO is how to demonstrate that the organisation is accountable and can demonstrate GDPR compliance.

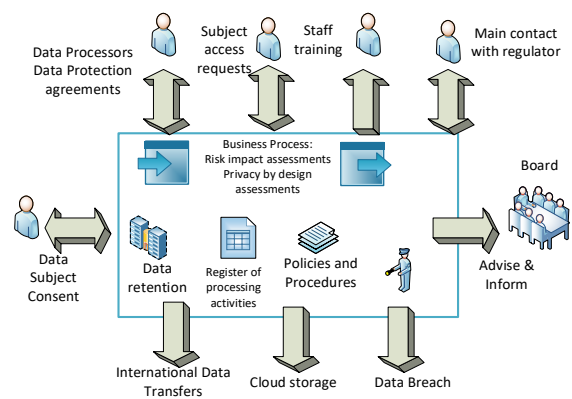

Figure 2: The breadth and complexity of the role of Data Protection Officer (Source Author).

## 3 CURRENT APPROACHES TO GDPR COMPLIANCE

This section discusses the broad range of tools and methods that are available to DPO's to demonstrate the GDPR compliance of their organisation.

## 3.1 Private Enterprise Software Solutions

There has been a call for tools and methods to assist organisations in meeting their GDPR compliance obligations (Piras, 2019). This is being met by large financial investments by venture capital companies with over $500 million invested in privacy related start-ups around the world in 2017 (IAPP, 2019) There are over 263 vendors offering privacy software tools to organisations (IAPP, 2019). These software solutions come in many forms ranging from simple questionnaires and templates to solutions that focus on individual aspects of compliance for GDPR such as website scanning for use of cookies. The main categories of these privacy tools are as follows (IAPP, 2019):
- Activity Management – control and monitor access to personal data
- Assessment Managers - automate different functions of a privacy program, locating risk gaps, demonstrating compliance
- Consent managers - help organizations collect, track, demonstrate and manage users' consent.
- Data discovery – determine and identify personal data held
- Data mapping solutions - determine data flows throughout the enterprise.
- De-identification pseudonymisation tools
- Secure Internal Enterprise communications
- Data Breach Incident response solutions
- Privacy information managers - provide latest privacy laws around the world.
- Website scanning – catalogue cookies

Table 1: Privacy software tools, number of vendors per category – (IAPP 2019).

| Privacy Product Category | No. of Vendors offering this service |
|---|---|
| Activity Monitoring | 86 |
| Assessment Manager | 105 |
| Consent Manager | 82 |
| Data Discovery | 94 |
| Data Mapping | 117 |
| De Identification/Pseudonymity | 46 |
| Enterprise Communications | 39 |
| Incident Response | 63 |
| Privacy Information Manager | 73 |
| Website Scanning | 30 |

Whilst there are a variety of privacy software solutions being offered by vendors, as displayed in Table 1 "there is no single vendor that will automatically make an organization GDPR compliant" (IAPP 2018). In fact, most solutions on offer from private enterprise cover 3 or less categories, see Figure 3.

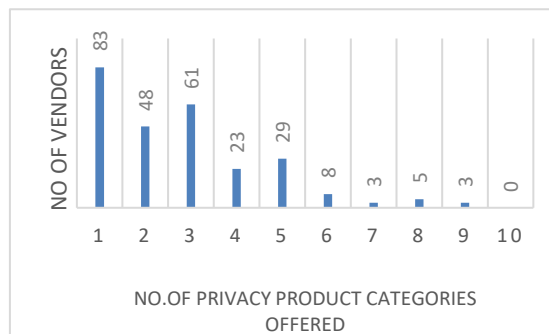

Figure 3: No. of privacy product categories offered by no. of vendors.

An accountability framework requires a comprehensive approach to compliance across the organisation. Whilst these software solutions go some way towards the demonstration of compliance, the author has identified several weaknesses in these private enterprise software solutions, as follows:
- They are not supported by published methodologies or evidence to support their validity or even utility
- Many of these solutions are stand - alone in that they lack inter-operability with other GDPR compliance systems and hence cannot easily be assembled into toolchains providing comprehensive compliance reports and metrics, quality improvement processes or data analytics such as root cause analysis
- They focus on manual or semi-automated assessment approaches that are labour intensive, rely on domain experts and are not driven by quantitative operational data that is increasingly being generated by organisations
- They are created by private enterprise and are based on an interpretation of the regulation, rather than being developed with the input of the regulator.

These solutions offer a starting point for GDPR compliance for an organisation however the lack of academic rigour or formal regulatory input and the inability to connect and build tool chains inhibits these solutions. The use of data driven inputs from heterogeneous sources and the mapping of business processes using agreed semantic standards would improve inputs to the evaluation tool. This would remove subjectivity and improve the quality of the outputs. GDPR compliance software must avoid the

"pitfalls of a fragmented Tower of Babel approach." (Butler, 2018). The best of breed software point solution products could be used to feed a global evaluation tool to optimise and organise the outputs using agreed semantics.

## 3.2 Maturity / Capability Models

Capability Maturity Models have been used for compliance monitoring for many years (Humphrey, 2002). The American Institute of Certified Public Accountants privacy maturity model (AICPA, 2011) was used to gain an understanding of an organisation's privacy compliance standing. It used a set of questions referred to as "generally accepted privacy principles" in the form of 73 measurable criteria. It gauged compliance along an axis of five maturity levels from ad hoc to optimized. The drawbacks of this methodology as a measure of compliance to the GDPR are that it predates the GDPR and would therefore need updating to reflect the new regulation. The more recent IAPP Maturity Framework (2019) develops a series of checklists built through "collaboration between a team of highly experienced privacy and security professionals, lawyers and regulators." Both solutions provide visualisations of compliance on an axis and are an indicative measure of compliance. However, they do have a number of drawbacks as follows:
- They are labour intensive and dependant on highly skilled labour/domain experts
- They are prone to human subjectivity, bias and errors
- They are infrequently updated
- The measures chosen utilise a selection of questions and checklists that require academic validation
- They are not suitable as part of an automated process and quality improvement toolchain

Whilst these maturity models are indicative of an organisations GDPR compliance position, the limitations outlined prevent these tools from developing any further without automation. Once automated, the lack of reporting and interoperability standards mentioned in the last section become relevant.

## 3.3 Self-assessment Checklists from Regulatory Authorities

Several data protection supervisory authorities have provided self-assessment checklists and accountability toolkits to assist organisations to prepare for GDPR. These come in the form of a series of questions and check-lists and are designed to assist the organisation in checking their compliance level. These toolkits are devised to provide broad coverage of all the principles of the GDPR. Just like maturity models these checklists provide an overview of compliance, however the main drawbacks of these tools for GDPR compliance are that they are fundamentally high-level self-assessments tools and are generic by nature and lack depth. Like maturity models, they rely on qualitative input of users, and they lack input or output interoperability with other solutions. However, the key benefit of these checklists and toolkits are that they have been developed by regulators, unlike maturity models and private enterprise software solutions, which have been developed independently.

## 4 CHALLENGES FOR THE NEXT GENERATION OF COMPLIANCE TOOLS – LESSONS FROM RegTech

In this section this paper will look at the emergence of RegTech as a solution to compliance challenges in the financial industry. RegTech can be defined as "the use of technological solutions to facilitate compliance with, and monitoring of regulatory requirements" (Colaert, 2017). The financial crisis of 2008 brought about a significant increase in new compliance legislation. (Butler, 2019). The emergence of RegTech came about for the following reasons (Arner, 2017):
- Enhanced compliance requirements
- Developments in data science and Artificial Intelligence
- Cost of compliance
- Regulators efforts to enhance the efficiency of supervisory tools

The key drivers for the RegTech technological solutions have been to make compliance reporting simple, easy and efficient and to reduce the risks of individual errors or liability and to build automated systems to facilitate legal compliance. RegTech has the potential to enable organisations to use business data to enhance better decision making and quickly identify non- compliances (Butler, 2019).

When we look at RegTech solutions we see compliance technology software spanning a wide breadth from its simplest form such as automated reporting or dashboard views to complex tools for carrying out specific regulatory functions (Colaert, 2017). Some examples of RegTech solutions are in the area of anti-money laundering where large financial deposits can be automatically detected and reported to a compliance officer, thus reducing the risk of human error in the form of an inattentive staff member. Similarly, Markets in Financial Instruments (MiFID) tests, help organisations to determine what level of investment advice must be given to a customer based on the results of an automatically processed questionnaire (Colaert, 2017). Again, this solution helps an organisation to reduce errors and meet its legal obligations through process automation. These solutions remove the need for human intervention and make compliance less complex. RegTech tools are being used to leverage data from existing operational information systems and seek to provide agile solutions to improve compliance visibility, through the automation of mundane compliance tasks and reduce risk to the organisation (Colaert, 2017).

The foundation of compliance has been to prevent identify, respond to and remedy risk. (Deloitte, 2016). RegTech solutions are being developed to meet these regulatory requirements, but also to modernise compliance and generate a measurable value proposition to the organisation. RegTech solutions enhance the basics of compliance through enhanced data integration, the use of automation, predictive analytics and strategic process alignment (Deloitte, 2016).

The role of the supervisory authority has arguably been transformed by RegTech (Arner, 2017). The regulator not only has access to periodic or real-time, fine-grained compliance reports, and the incremental improvements in compliance but they are promoting the design of a regulatory framework able to dynamically adapt to new rules and regulations (Arner, 2017).

The contrast between the innovation in this space and the GDPR compliance tools discussed above suggests that the use of a RegTech approach applied to the GDPR would yield significant benefits to DPO's, organisations and regulators. It may even side-step the crisis in GDPR certification schemes by providing automated transparent accountability that regulators can query and analyse without recourse to a slow third-party certification service. This blend of technology can yield significant benefits for organisations (Arner, 2017).

# 5 REQUIREMENTS FOR GDPR RegTech

In this section, this paper takes the learning from RegTech as described in section four and proposes a RegTech approach to GDPR compliance. This design takes it's learning from RegTech where common design protocols and agreed semantic standards (Butler, 2019) are used to integrate new heterogenous tools to provide the organisation with the necessary information to monitor, evaluate and report compliance (See Fig. 4). This approach allows for new tools to be integrated seamlessly. The RegTech approach to compliance seeks to automate data inputs to reduce human errors and remove subjectivity. The use of common standards, protocols and semantics facilitates a flexible, nimble and agile and cost-effective approach to compliance. The next generation of GDPR compliance tools need to consider a RegTech approach to meet their accountability obligations.

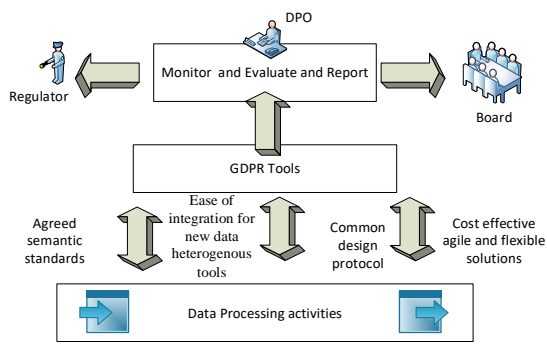

Figure 4: A RegTech approach to GDPR compliance (Source Author).

# 6 PROOF OF CONCEPT – AN EVALUATION TOOL FOR GDPR COMPLIANCE

In this section we describe a prototype GDPR high-level evaluation tool that has been developed based upon the developments in RegTech, outlined in section 5. The tool is an open-source high level GDPR compliance evaluation methodology that has been based on a self-assessment checklist created by a data protection regulator. It has been developed to measure the GDPR compliance level in an organisation. The evaluation tool was developed from the Irish Data Protection Commission self-assessment checklist which segmented the GDPR into 8 regulatory sections and posed 54 questions in total. The tool is designed to be a layered information delivery system that provides information and insights so that the DPO can measure, monitor and manage business performance more effectively, and address accordingly (Eckerson, 2010).

The evaluation tool provides three layers of data as displayed in Fig. 5. The top level being a graphical overview of compliance for monitoring and reporting purposes, the second layer being the dimensional data that provides a view of each aspect of the GDPR and the final layer that being the detail of each GDPR compliance area.

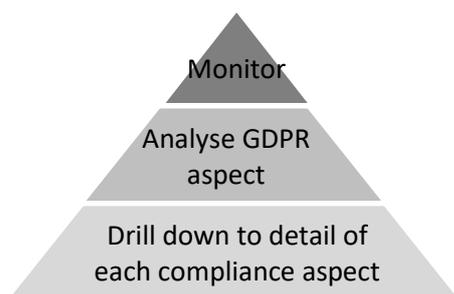

Figure 5: GDPR Evaluation dashboard overview (Eckerson, 2010).

It is planned that the tool will use the W3C Community group's data protection vocabulary (Pandit, 2019) to describe the context using explicit semantics and the W3C Data Cube vocabulary to represent the time series of measurements across the different GDPR aspects or dimensions (Cyganiak, 2014). This development involved taking the self-assessment checklist and transforming it into an evaluation tool which was populated by a sample organisation each month for six months in total. The overall GDPR compliance monthly score for the organisation for each month is displayed in figure 6. This information gives the DPO a high-level view of compliance for the organisation.

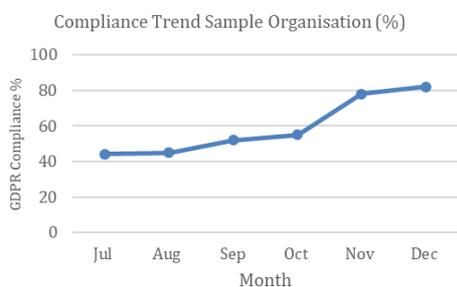

Figure 6: Compliance Trend for Sample organisation.

The results from the evaluation tool can be also viewed by GDPR regulatory section to analyse how the organisation is performing in the various aspects of GDPR compliance, thus providing enhanced visibility to the DPO. In table 2 the organisation is fully compliant in accuracy and retention but is only 50% compliant regarding data breaches. The data can be examined to another sub-level to provide the detail by GDPR aspect. Table 2 breaks out the Data Breach aspect and provides the granularity that a DPO needs to provide feedback to the controller to drive actions and improve the compliance of the organisation.

Table 2: Compliance score per regulatory area.

| GDPR Section | Compliant % |
|---|---|
| Personal data | 67% |
| Data subject rights | 40% |
| Accuracy and retention | 100% |
| Transparency requirements | 100% |
| Other data controller obligations | 83% |
| Data security | 100% |
| Data breach | 50% |
| International data transfers | 100% |
| Total score | 82% |

In the sample organisation the DPO can identify the non-compliant areas as identified in table 3 and take the necessary actions to resolve.

Table 3: Non-compliance results for Data Breach.

| Data Breach section | Areas of failure |
|---|---|
| Are plans and procedures regularly reviewed? | non-compliant |
| Are all data breaches fully documented? | non-compliant |
| Are there cooperation procedures in place between data controllers, suppliers and other partners to deal with data breaches? | non-compliant |

This approach has demonstrated the use of a RegTech approach to GDPR compliance using a simple cost-effective method. It has utilised questions that have been created by regulatory authorities themselves so they could serve as a strong platform for the assessment of compliance. The evaluation tool meets the requirement of being comprehensive in that it covers the breadth of the GDPR and is informative in that it provides specific scores into GDPR areas requiring focus. The evaluation process is repeatable in that it can be run at intervals to generate compliance trends. The results yielded specific and relevant scores that can be used to drive corrective actions. The use of data driven inputs from heterogeneous sources and the mapping of business processes into the evaluation tool using agreed semantic standards would remove qualitative user inputs and would improve inputs to the evaluation tool. This would remove subjectivity and improve the quality of the outputs.

# 7 CONCLUSIONS

Organisations are accountable for the demonstration of their compliance with the GDPR regulation. We have seen that the available compliance tools go some way to achieving this goal, but each have their shortcomings. A RegTech approach to GDPR compliance has shown that the use of technology to improve compliance monitoring and reporting can be achieved when flexible, agile, cost effective, extensible and informative tools are combined. The opportunities to further develop GDPR compliance tools exists if agreed semantic standards (Butler, 2019) are developed to automate processes and remove subjectivity from data inputs. We conducted a proof of concept to demonstrate the application of some of these RegTech approaches to GDPR Compliance. A GDPR compliance tool was developed to monitor and analyse organisational compliance that yielded a GDPR compliance output for an organisation. The compliance report that was generated from the evaluation tool can be used to identify GDPR areas where the organisation is not compliant, to trend their progress towards GDPR compliance over time and to benchmark performance versus other organisations. The DPO can use the results to direct resources to areas of non-compliance and improve their score, thus reducing the risk of GDPR fines. We have shown that a RegTech approach to GDPR can enable an organisation to meet its obligations to comply with the accountability principle.

# ACKNOWLEDGEMENTS

This work is partially supported by Uniphar PLC., and the ADAPT Centre for Digital Content Technology which is funded under the SFI Research Centres Programme (Grant 13/RC/2106) and is co-funded under the European Regional Development Fund.